\documentclass[runningheads,a4paper]{llncs}

\usepackage{amssymb}
\usepackage{graphicx}
\newtheorem{defn}{Definition}
\usepackage{threeparttable}
\usepackage{multirow}
\usepackage{array}

\usepackage{url}
\urldef{\mailsa}\path|{alfred.hofmann, ursula.barth, ingrid.haas, frank.holzwarth,|
\urldef{\mailsb}\path|anna.kramer, leonie.kunz, christine.reiss, nicole.sator,|
\urldef{\mailsc}\path|erika.siebert-cole, peter.strasser, lncs}@springer.com|    
\newcommand{\keywords}[1]{\par\addvspace\baselineskip
\noindent\keywordname\enspace\ignorespaces#1}

\begin{document}

\mainmatter  

\title{CompetitiveBike: Competitive Prediction of Bike-Sharing Apps Using Heterogeneous Crowdsourced Data}

\titlerunning{CompetitiveBike}

%
%
\author{Yi Ouyang\inst{1}%
\and Bin Guo\inst{1}
\and Xinjiang Lu\inst{1}
\and Qi Han\inst{2}
\and\\
Tong Guo\inst{1}
\and Zhiwen Yu\inst{1}}
\authorrunning{Yi Ouyang et al.}

\institute{Northwestern Polytechnical University, Xi'an, Shaanxi, China\\
	\email{guob@nwpu.edu.cn}
	\and Colorado School of Mines, Golden, CO, USA}

%
%

\toctitle{Lecture Notes in Computer Science}
\tocauthor{Authors' Instructions}
\maketitle

\begin{abstract}
	In recent years, bike-sharing systems have been deployed in many cities, which provide an economical lifestyle. With the prevalence of bike-sharing systems, a lot of companies join the market, leading to increasingly fierce competition. To be competitive, bike-sharing companies and app developers need to make strategic decisions for mobile apps development. Therefore, it is significant to predict and compare the popularity of different bike-sharing apps. However, existing works mostly focus on predicting the popularity of a single app, the popularity contest among different apps has not been explored yet. In this paper, we aim to forecast the popularity contest between Mobike and Ofo, two most popular bike-sharing apps in China. We develop CompetitiveBike, a system to predict the popularity contest among bike-sharing apps. Moreover, we conduct experiments on real-world datasets collected from 11 app stores and Sina Weibo, and the experiments demonstrate the effectiveness of our approach.
\end{abstract}

\section{Introduction}

In recent years, shared transportation has grown tremendously, which provides us an economical lifestyle. Among the various forms of shared transportation, public bike-sharing systems \cite{demaio2009bike}, \cite{shaheen2010bikesharing}, \cite{pucher2010infrastructure} have been widely deployed in many metropolitan areas (e.g. New York City in the US and Beijing in China). A bike-sharing system provides short-term bike rental service with many bicycle stations distributed in a city \cite{liu2016rebalancing}. A user can rent a bike at a nearby bike station, and return it at another bike station near his/her destination. The worldwide prevalence of bike-sharing systems has inspired lots of active research, such as bike demand prediction \cite{chen2015bike}, \cite{chen2016dynamic}, \cite{liu2017functional}, bike rebalancing optimization \cite{singla2015incentivizing}, and bike lanes planning \cite{bao2017planning}.

More recently, station-less bicycle-sharing systems are becoming the mainstream in many big cities in China such as Beijing and Shanghai. Mobike\footnote[1]{\url{https://en.wikipedia.org/wiki/Mobike}} and Ofo\footnote[2]{\url{https://en.wikipedia.org/wiki/Ofo_(bike_sharing)}} are two most popular station-less bicycle-sharing systems. Unlike traditional public bike-sharing systems, station-less bike sharing systems aim to solve ``the last one mile'' issue for users. Using the Mobike/Ofo mobile app, users can search and unlock nearby bikes. When users arrive at their destinations, they do not have to return the bikes to the designated bike station. Instead, they can park the bicycles at a location more convenient for them. Therefore, it is easier for users to rent and return bikes than traditional bike-sharing systems.

As bike-sharing apps become increasingly popular, a lot of companies join the bike-sharing market, leading to fierce competition. To thrive in this competitive market, it is vital for bike-sharing companies and app developers to understand their competitors and then make strategic decisions accordingly \cite{xu2011mining} for mobile app development and evolution \cite{di2016would}. Therefore, it is significant and necessary to predict and compare the future popularity of different bike-sharing apps.

When users download and install a mobile app, they may submit user experience to the app store \cite{martin2017survey}, \cite{fu2013people}, \cite{gu2015parts}. Specifically, users may upload their requirements (e.g. functional requirements), preferences (e.g. UI preferences) or sentiment (e.g. positive, negative) through reviews, as well as their satisfaction level through ratings. Online social media is another way to share the user experience of a mobile app. When users actually use the bike, they may share the ride experience on social media. Specifically, users may record the feeling of the ride, the advantages and disadvantages of the bike/system, or the comparison with other bikes/systems. Both users' online and offline experience will affect the popularity of the apps, thereby affecting their popularity contest outcome. Therefore, app store data and microblogging data are complementary, and can describe a mobile app from different perspectives. In this paper, we study the problem of competitive prediction of bike-sharing apps using heterogeneous app store data and microblogging data. 

To the best of our knowledge, the problem of predicting the competitiveness of mobile apps has not been well investigated in the literature. There are several challenging questions to be answered. How to forecast the popularity contest outcomes of bike-sharing apps? How to extract effective features to characterize the competitiveness of bike-sharing apps from heterogeneous crowdsourced data?

To answer these questions, we propose CompetitiveBike, a system that predicts the outcomes of the popularity contest among bike-sharing apps leveraging heterogeneous app store data and microblogging data. We first obtain app descriptive statistics and sentiment information from app store data, and descriptive statistics and comparative information from microblogging data. Using these data, we extract both coarse-grained  and fine-grained competitive features. Finally, we train a regression model to predict the outcomes of popularity contest. We make the following contributions.

(1) This work is the first to study the problem of competitive prediction of bike-sharing apps. We use two indicators for the comparison: i) competitive relationship to indicate which app is more popular; and ii) competitive intensity to measure the popularity gap between the two apps/systems.

(2) To predict popularity contest between apps, we extract features from different perspectives including the descriptive information of apps, users' sentiment, and comparative opinions. Using the basic information, we further extract two novel features: coarse-grained and fine-grained competitive features, and choose Random Forest for prediction.

(3) To evaluate CompetitiveBike, we collect data about Mobike and Ofo from 11 app stores and Sina Weibo. With the data collected, we conduct extensive experiments from different perspectives. We find that the Random Forest model performs well on \emph{competitive relationship} prediction (the Accuracy is 71.4\%) as well as \emph{competitive intensity} prediction (the RMSE is 0.1886). A combination of the coarse-grained and fine-grained competitive features improves performance in popularity contest prediction, and a combination of data from app store and microblogging also improves performance in popularity contest prediction. The results demonstrate the effectiveness of our approach.

\section{Related Work}
\label{sec:relatedwork}

\subsection{App Popularity Prediction}
Recently, a significant effort has been spent on predicting popularity of mobile app \cite{zhu2015popularity}, \cite{malmi2014quality}, \cite{wang2017app}, \cite{finkelstein2013mining}. Zhu et al. \cite{zhu2015popularity} proposed the Popularity-based Hidden Markov Model (PHMM) to model the popularity information of mobile apps. Wang et al. \cite{wang2017app} proposed a hierarchical model to forecast the app downloads. Malmi \cite{malmi2014quality} found that there existed connection between app popularity and the past popularity of other apps from the same publisher. Finkelstein et al. \cite{finkelstein2013mining} found that there is a strong correlation between rating and the downloads.

Our work differs from and potentially outperforms the previous work in several aspects. First, we focus on the problem of competitive prediction of bike-sharing apps, instead of the prediction of a single app. Second, we predict the popularity contest leveraging heterogeneous crowdsourced data (i.e., app store data and microblogging data) that are often complementary and can reflect mobile app popularity contest from different perspectives.

\subsection{Competitive Analysis}
Competitive analysis involves the early identification of potential risks and opportunities to help managers making strategic decisions for an enterprise \cite{xu2011mining}. Jin et al. \cite{jin2016identifying} selected subjective sentences from reviews which discuss common features of competing products. He et al. \cite{he2013social} analyzed the textual content on the social media of the three largest pizza chains, and the results revealed the business value of comparing social media content. Maksim et al. \cite{tkachenko2016comparative} proposed a generative model for comparative sentences, jointly modeling two levels of comparative relations: the level of sentences and the level of entity pairs. Zhang et al. \cite{zhang2013product} proposed to scan reviews to update a product comparison network. 

These studies conduct competitive analysis simply via semantic analysis of users' opinion. In contrast, our work extracts features from different perspectives including the descriptive information of apps, user's sentiment, and comparative opinions. Using the basic information, we further extract coarse-grained and fine-grained competitive features, and train a model to predict popularity contest.

\section{Data Acquisition and Analysis}

\subsection{App Store Data}
We collected data from 11 mainstream Android app stores\footnote{Data from Google Play is more sparse than these app stores as Mobike and Ofo users are mainly from China, so we did not collect data from Google Play.} in China, including: Wandoujia, Huawei, 360, Meizu, OPPO, VIVO, Yingyongbao, Xiaomi, Baidu, Lenovo and Anzhi market. An overview of app store data is listed in Table~\ref{tab:appstatistics}. 

\begin{table}
	\caption{Basic Statistics of the App Store Data}
	\label{tab:appstatistics}
	\centering
	\begin{tabular}{lr}
		\hline
		Property & Statistics\\
		\hline
		App stores & 11\\
		Time span & 04/22/2016 - 03/14/2017\\
		Reviews of Mobike & 69,228\\
		Reviews of Ofo & 13,928\\
		Total downloads of Mobike & 35,591,757\\
		Total downloads of Ofo & 30,423,077\\
		\hline
	\end{tabular}
\end{table}%

We collected data between 04/22/2016 and 03/14/2017. At the beginning, these two apps were still relatively new and they are not as popular now, so there were not a lot of data. To ensure prediction accuracy, the actual time span of the app store data we use is from 06/20/2016 to 03/12/2017, exactly 38 weeks. 

\begin{figure}
	\centering
	\includegraphics[height=35mm]{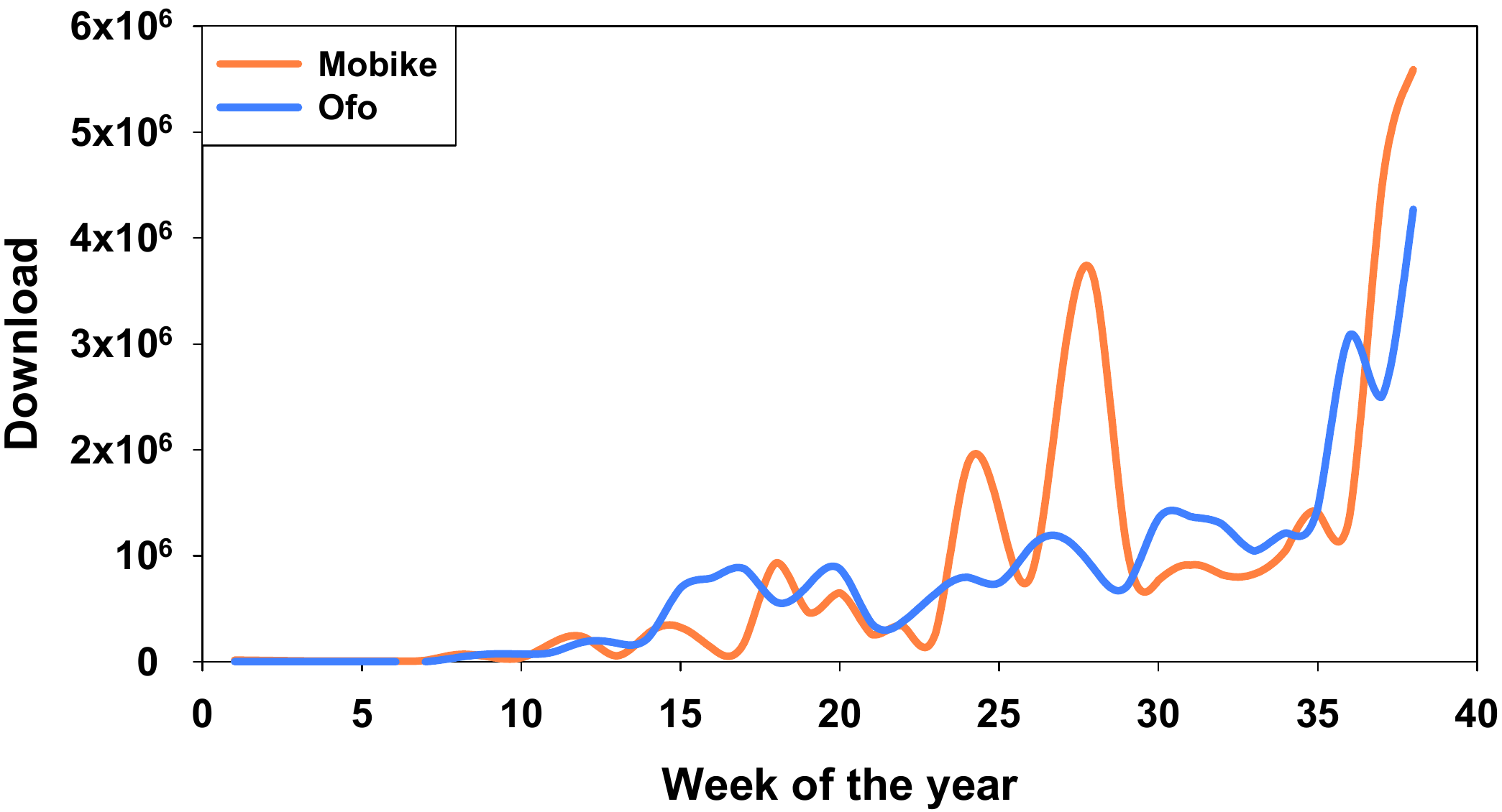}
	\caption{The weekly downloads of Mobike and Ofo.}
	\label{fig:weekly_download}
\end{figure}

Figure~\ref{fig:weekly_download} shows the weekly downloads of the two apps. We can observe that their downloads are all increasing, and for the recent months, Mobike and Ofo have comparable downloads.

\subsection{Microblogging Data}
We crawled three microblogging datasets from Sina Weibo\footnote{\url{https://weibo.com/}}, the most popular microblogging service in China. The first dataset was crawled by using a combination of the two keywords ``Mobike'' and ``Ofo'', we refer it as the ``Mobike \& Ofo''. The second one was crawled by using the keyword ``Mobike'', we refer it as the ``Mobike''. The third one was crawled by using the keyword ``Ofo'', we refer it as the ``Ofo''. An overview of three datasets is listed in Table~\ref{tab:microblog}.

\begin{table}
	\caption{Basic Statistics of the Microblogging Data}
	\label{tab:microblog}
	\centering
	\begin{tabular}{ccccccc}
		\hline
		Dataset & Time Span & Microblogs & Users & Reposts & Comments & Likes\\
		\hline
		Mobike \& Ofo & 06/21/2016 - 03/14/2017 & 11,176 & 8,725 & 34,801 & 35,646 & 31,295\\
		Mobike & 04/22/2016 - 03/14/2017 & 52,718 & 40,187 & 151,126 & 207,926 & 181,560\\
		Ofo & 05/30/2016 - 03/14/2017 & 43,746 & 35,752 & 145,882 & 181,815 & 170,644\\						
		\hline
	\end{tabular}
\end{table}%

\section{Problem Statement and System Framework}

\subsection{Problem Statement}
The problem can be stated as follows: given the app store data and microblogging data about Mobike and Ofo, we want to predict which app will be more popular in the future.

\begin{defn}
	\textbf{Popularity Contest}. \rm Inspired by \cite{desarbo2006competes}, the popularity of Mobike (or Ofo) can be measured by the downloads, and the popularity contest ($PC$) between Mobike and Ofo can be defined by the difference in their downloads $D_m$ and $D_o$:
\end{defn}

\begin{equation}
\label{eqn:1}
PC=\frac{D_m-D_o}{D_m+D_o}
\end{equation} 

\begin{defn}
	\textbf{Competitive Relationship}. \rm The competitive relationship ($CR$) between Mobike and Ofo can be one of the two possbilities: 1) Mobike is more popular than Ofo, or 2) Ofo is more popular than Mobike.  According to Formula~(\ref{eqn:1}), when $PC>0$, Mobike is more popular; otherwise, Ofo is more popular.
\end{defn}

\begin{defn}
	\textbf{Competitive Intensity}. \rm The competitive intensity ($CI$) between Mobike and Ofo is the absolute value of $PC$. The smaller the value, the higher the competitive intensity is. 
\end{defn}

Formally, we extract feature set $X$ from app store data and microblogging data, then we want to predict the popularity contest $Y$. Let $X=\{x_1,...x_N\}$ and $Y=\{y_1,...y_N\}$, given $X^{(1:t+1)}(=\{X^{(1)},...,X^{(t+1)}\})$ and $Y^{(1:t)}(=\{Y^{(1)},...,Y^{(t)}\})$, our objective is to predict $Y^{(t+1)}$. 

\subsection{System Framework}
\begin{figure}
	\centering
	\includegraphics[width=70mm]{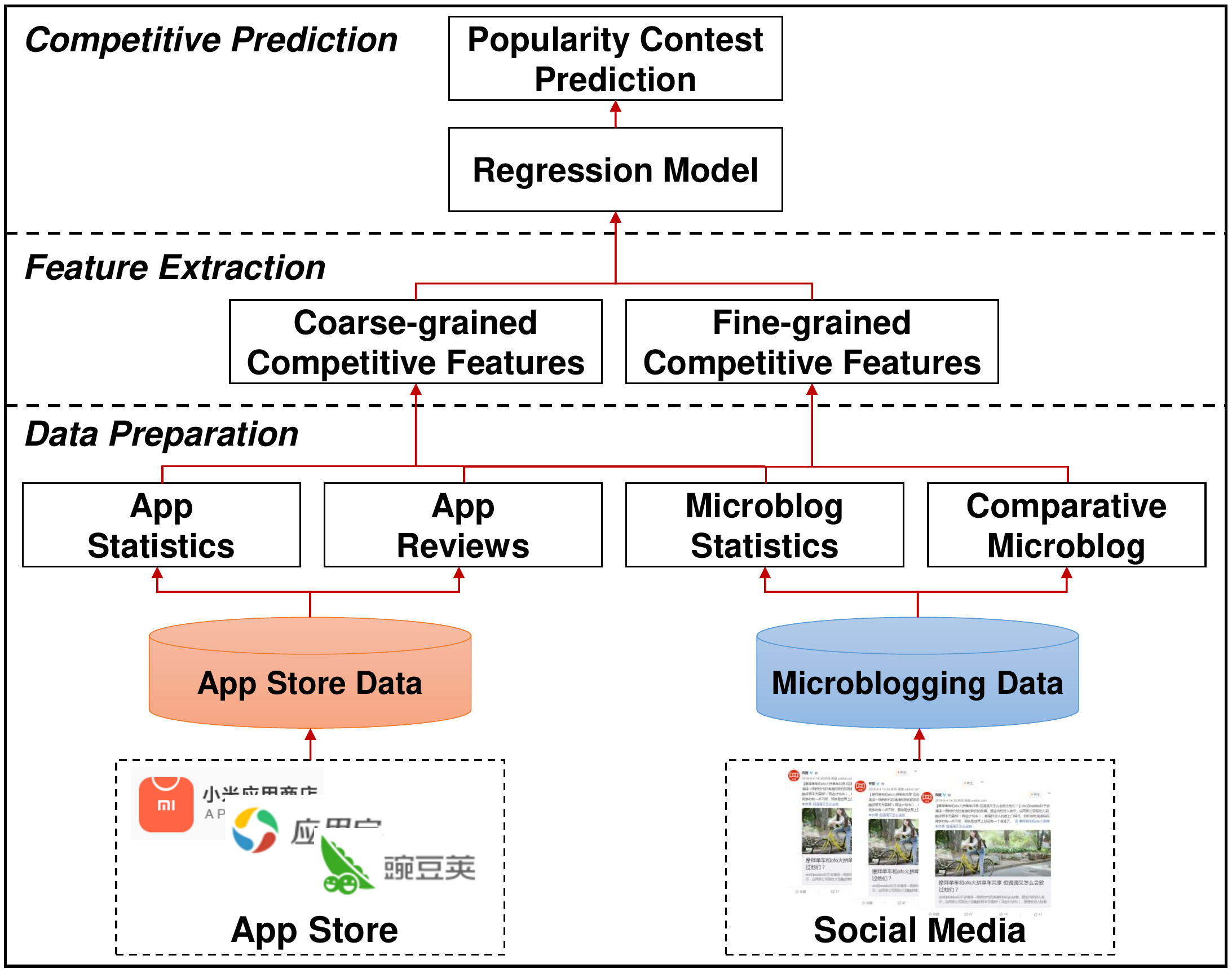}
	\caption{System framework.}
	\label{fig:framework}
\end{figure}

The overview of the framework is illustrated in Figure~\ref{fig:framework}, which mainly consists of three layers: data preparation, feature extraction, and competitive prediction.

\emph{\textbf{Data Preparation.}} We obtain app statistics and reviewers' sentiment from app store data, and microblogging statistics and comparative information from microblogging data.

\emph{\textbf{Feature Extraction.}} To effectively extract and quantify the factors impacting mobile app popularity contest, we extract features from different perspectives including the inherent descriptive information of apps, users' sentiment, and comparative opinions. With this information, we further extract two novel sets of features: coarse-grained and fine-grained competitive features.

\emph{\textbf{Competitive Prediction.}} With these two extracted feature sets, we train a model to predict the popularity contest between Mobike and Ofo.

\section{Popularity Contest Prediction}
In this section, we first analyze the factors impacting the popularity contest between Mobike and Ofo, then extract coarse-grained and fine-grained competitive features from these factors to characterize popularity contest. Finally, we train a model to predict popularity contest.

\subsection{Coarse-grained Competitive Features}

\subsubsection{Features from App Store.}
When users download and install a mobile app, they may submit reviews and ratings to the app store. For example, a user wrote: ``\emph{The Mobike app cannot launch today, it was still okay yesterday, what's the matter? It's terrible!}'' According to the review, we believe that app store data (e.g. reviews, ratings) can reflect users' online experience with the app. Typically, users may upload their requirements (e.g. functional requirements), preferences (e.g. UI preferences), or sentiment (e.g. positive, negative) through reviews, and they may also rate the app based on their overall satisfaction. Therefore, we extract features from reviews and ratings to characterize popularity contest.

\emph{\textbf{App Statistics.}} Generally, the numerical statistics of reviews and ratings in each time window can reflect the popularity of the app. In other words, a bigger number of reviews and a higher rating score may indicate that the app is more popular. We use the difference between app's review number $DN$ (and rating scores $DS$) to characterize popularity contest. A small value of $DN$ (and $DS$) indicates that they have similar number of reviews (and rating score), thus their competition is more intense.

\emph{\textbf{Sentiment Similarity.}} Besides numerical statistics, app reviews can express users' sentiment. We use a Chinese sentiment analyzer called SnowNLP\footnote{\url{https://github.com/isnowfy/snownlp}} to analyze the sentiment of reviews. We calculate the sentiment value $s_i$ of each review at time instant $t_i$,  then we obtain the sentiment distribution vector $\mathbf{v_i}=(p_1, p_2, p_3)$ at time $t_i$, where $p_1$, $p_2$, $p_3$ is corresponds to negative, neutral and positive sentiment proportion respectively.

The extracted sentiment sequences are only for a single app, when we consider the competition between two apps, we compute sentiment similarity to capture the difference of users' sentiment about these apps, and the similarity can be measured by calculating the cosine similarity \cite{ouyang2016sentistory}. The higher similarity means that users' opinions about them are more similar, and the competition between them is more intense. 

\subsubsection{Features from Microblogging.}
When users ride the bike of different apps, they may share their riding experience on social media. An example of a microblog is like this: ``\emph{This is my first ride of Mobike, it is so cool!}'' We believe online social media is another way to express users' riding experience. Therefore, we extract features from microblogging data to help understand the popularity contest of different apps.

\emph{\textbf{Microblogging Statistics.}} In the ``Mobike \& Ofo'' dataset, the number of microblogs, users, reposts, comments, and likes can reflect the attention about Mobike and Ofo on microblogging, the bigger value indicates more intense competition between Mobike and Ofo. 

In the ``Mobike'' dataset, more microblogs that contain the keyword ``Ofo'' imply that Ofo is more frequently mentioned in the ``Mobike'' dataset.  We use the ratio ($R_{om}$) of ``Ofo'' and ``Mobike'' to characterize the competition. Formally, $R_{om}=\frac{MN_o}{MN_m}$, where $MN_o$ and $MN_m$ represent the number of microblog that contains ``Ofo'' and ``Mobike'', respectively. Similarly, in the ``Ofo'' dataset, we  use the ratio ($R_{mo}$) of ``Mobike'' and ``Ofo'' to characterize the competition. The higher ratios, the more intense competition.

\emph{\textbf{Comparative Analysis.}} In addition to the numerical statistics, the textual information in microblog content is also valuable. The ``Mobike \& Ofo'' dataset often contains the comparison between Mobike and Ofo. Let us consider a microblog: ``\emph{Mobike is too heavy, and it is uncomfortable to ride. It is also slightly expensive. Of course, there are some aspects where Mobike is better than Ofo, such as: Mobike is more solid than Ofo, and its bell is also better.}'' According to this post, we observe that (1) there exists comparison between Mobike and Ofo; (2) a single microblog may compare the apps many times on different aspects (e.g. price, quality); (3) each comparison can discuss the advantages and disadvantages of the bike. Therefore, we need to address three issues in comparative analysis: (1) how to identify comparison between Mobike and Ofo; (2) how to calculate the comparison count; (3) how to determine the {\it comparison direction}, which means whether Mobike is better than Ofo, or Ofo is better than Mibike. We next describe our methods to address these issues. 

First, the occurrences of comparative words such as ``better'' often indicate  comparison and these comparative words are usually adjective or adverb. Therefore, to identify the comparison, we try to determine whether there exist comparative words in microblogs. Specifically, we use a Chinese lexical analyzer called Jieba\footnote{\url{https://github.com/fxsjy/jieba}} to annotate part of speech, and extract adjectives and adverbs to build a dictionary. We then determine whether there exist comparative words by querying the dictionary and filtering out microblogs without comparative words. After this, all the remaining microblogs contain comparison between Mobike and Ofo.

Next, when calculating the comparison count, we do not need to differentiate which aspects are in comparison. We can count the number of comparative word to determine the comparison count.

Last, the sentiment of the comparative words can be used to infer comparison direction. In the example above, ``Mobike is more solid than Ofo'' implies that Mobike is better than Ofo. We divide the dictionary into two sub-dictionaries: positive and negative. With a positive comparative word, 1 is added to its own score; with a negative comparative word, 1 is added to the score of the competitor. This way, we can obtain the comparison direction scores for Mobike and Ofo. We use the scores to characterize popularity contest.

\subsection{Fine-grained Competitive Features}
Each coarse-grained competitive feature is a time series with time window of one week. In each time window, we extract the temporal dynamics of the coarse-grained competitive features as the fine-grained competitive features to characterize the trend of the sequence \cite{lu2016characterizing}.

\emph{\textbf{Overall Descriptive Statistics}} describe the basic properties of the coarse-grained competitive features from multiple aspects. We extract the mean, standard deviation, median, minimum and maximum as features.

\emph{\textbf{Hopping Counts}} can effectively describe the ``pulse" of sequence and is calculated as the number of elements whose values are greater than their next element. This feature is used to characterize the fluctuation of the sequences. 

\emph{\textbf{Lengths of Longest Monotonous Subsequences}} describe the size of gradient descent or ascent patterns in a sequence. We examine the longest monotone (including increasing and decreasing) subsequences, and use the lengths of these two subsequences to describe the tendency of the sequence.

\subsection{Popularity Contest Prediction}
With these two extracted feature sets, we want to predict the popularity contest in the future, we use regression-based methods. Since the extracted features are sequences, and the time window is one week, we treat successive several weeks as the training set, then compare the state-of-the-art regression models. Section~\ref{sec:7} has the details on the models we compared and the one we eventually use.

\section{Performance Evaluation}
\label{sec:7}

\subsection{Experimental Setup}

\subsubsection{Comparison Settings.}
To demonstrate the effectiveness of different types of features, we divide the extracted features into two categories: (1) \emph{coarse-grained competitive features} (CF); (2) \emph{fine-grained competitive features} (FF). 

To demonstrate the effectiveness of heterogeneous crowdsourced data, we divide the features into another two categories according to the data source: (1) \emph{features from app store data} (AF); (2) \emph{features from microblogging data} (MF).

Regarding algorithm comparison, in the phase of \emph{competitive relationship} prediction, we evaluate three state-of-the-art classification algorithms: Decision Tree (DT), Adaboost and Random Forest (RF). In the phase of \emph{competitive intensity} prediction, we evaluate two state-of-the-art regression algorithms: Support Vector Regression (SVR) and Random Forest (RF).

To conduct popularity contest prediction, we use the following setup: we use ten successive weeks as the training set and the next one week as the test set.

\subsubsection{Baseline Algorithms.}
For popularity contest prediction, we use the following methods as the baselines:

\begin{itemize}
	\item \emph{Last\_predcition}: it predicts the popularity contest using the last one week, i.e. $Y^{(t+1)}=Y^{t}$. We refer it as ``Last''.
	\item \emph{CF}: it predicts the popularity contest using the coarse-grained competitive features alone.
	\item \emph{FF}: it predicts the popularity contest using the fine-grained competitive features alone.
	\item \emph{AF}: it predicts the popularity contest using the features from app store alone.
	\item \emph{MF}: it predicts the popularity contest using the features from microblogging platform alone.
\end{itemize}

\subsubsection{Evaluation Metrics.}
For popularity contest prediction, we measure the prediction performance using the following metrics:

\begin{itemize}
	\item In the phase of \emph{competitive relationship} prediction, we use \emph{Accuracy}, \emph{Precision}, \emph{Recall}, \emph{F-measure} as the evaluation metrics. Higher values of these metrics means the better performance in \emph{competitive relationship} prediction.
	\item In the phase of \emph{competitive intensity} prediction, we use \emph{RMSE} as the evaluation metric. A smaller RMSE means the better performance in \emph{competitive intensity} prediction.
\end{itemize}

\subsection{Experimental Results}

\subsubsection{Comparison of Different Algorithms.}
We want to compare the effectiveness of different algorithms in popularity contest: \emph{competitive relationship} and \emph{competitive intensity}. 

Regarding the \emph{competitive relationship} prediction, Figure~\ref{fig:algorithms} shows the Accuracy, Precision, Recall and F-measure of DT, Adaboost and RF. We observe that RF outperforms the other algorithms, with the Accuracy of 71.4\%, and the state-of-the-art classification algorithms outperforms the baselines.

Regarding the \emph{competitive intensity} prediction, Table~\ref{tab:algorithms} shows the RMSE of Last, SVR, and RF. We observe that RF again outperforms other algorithms, and the RMSE of the baseline is much larger than RF regression algorithm. 

In summary, the state-of-the-art machine learning algorithms can train a better learning model by using the proposed features. RF performs well on \emph{competitive relationship} prediction as well as \emph{competitive intensity} prediction. Therefore, we choose RF as the default predictor for predicting popularity contest.

\vspace{0.3cm}
\makeatletter\def\@captype{figure}\makeatother
\begin{minipage}{0.45\linewidth}
	\centering
	\includegraphics[height=40mm]{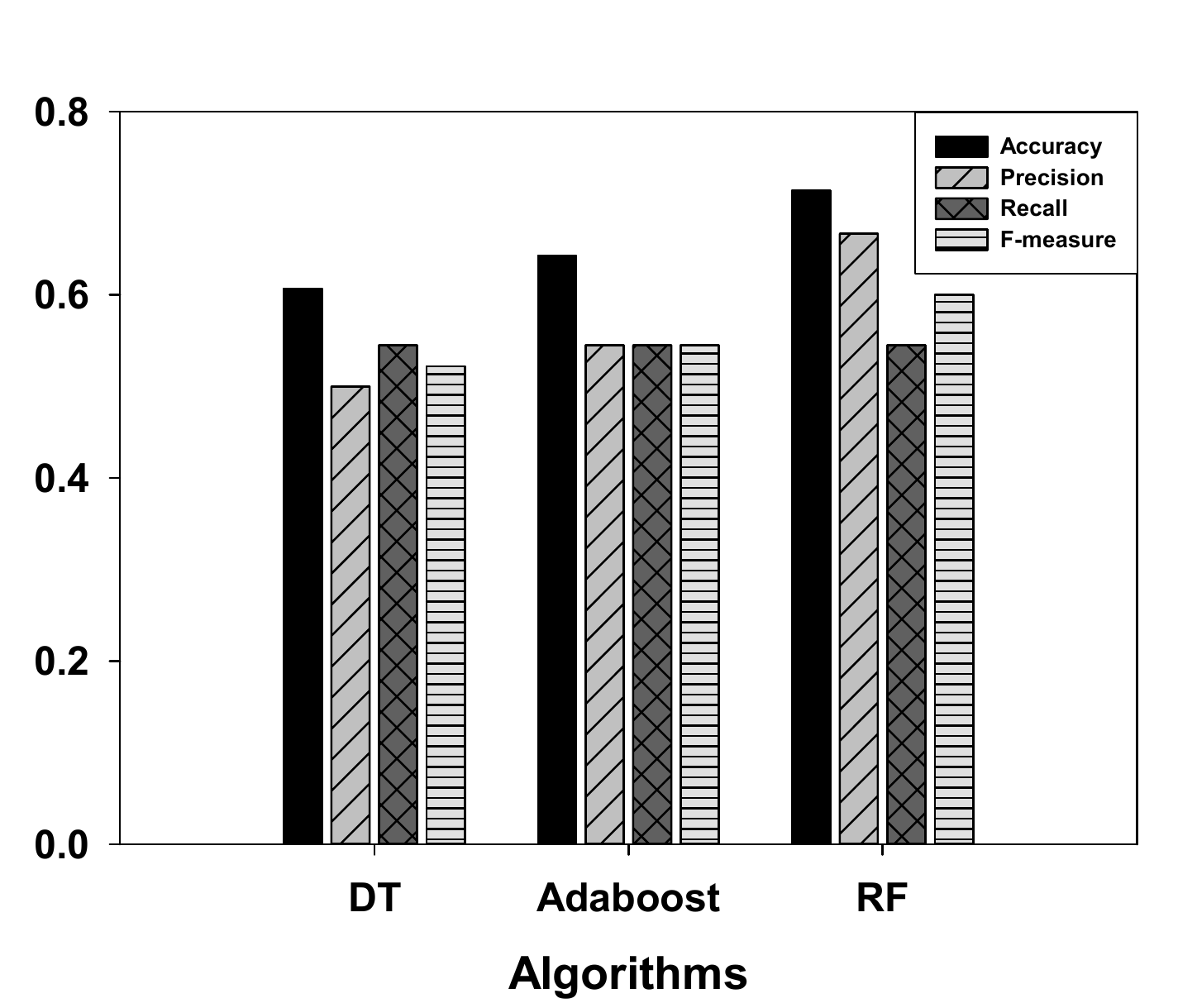}
	\caption{Comparison of Algorithms.}
	\label{fig:algorithms}
\end{minipage}
\makeatletter\def\@captype{table}\makeatother
\begin{minipage}{0.55\linewidth}
	\centering
	\caption{RMSE of Algorithms}
	\label{tab:algorithms}
	\vspace{0.3cm}	
	\begin{tabular}{m{1.5cm}<{\centering} m{1.5cm}<{\centering} m{1.5cm}<{\centering}}
		\hline
		Last & SVR & RF\\
		\hline
		0.2280 & 0.2185 & 0.1886\\
		\hline
	\end{tabular}
\end{minipage}

\subsubsection{Comparison of Different Features.}
We try to determine whether the combination of the coarse-grained  and fine-grained competitive features can improve the performance of prediction. Therefore, we compare the CF, FF, and CF+FF, respectively.

Figure~\ref{fig:features} shows the Accuracy, Precision, Recall and F-measure of CF, FF and CF+FF. We observe that FF outperforms CF, with the Accuracy of 67.9\%, while CF is 60.7\%. This is because FF is generated based on CF, and it can reflect the fine-grained tendency of CF. Furthermore, the combination of the coarse-grained and fine-grained competitive features (CF+FF) improves the performance in \emph{competitive relationship} prediction, compared with CF and FF alone.

Table~\ref{tab:features} shows the RMSE of CF, FF and CF+FF. We can observe that FF outperforms CF, and can reflect the temporal dynamics of the CF. Furthermore, the combination of the coarse-grained and fine-grained competitive features (CF+FF) improves the performance in \emph{competitive intensity} prediction, compared with CF and FF alone.

In summary, FF outperforms CF in both \emph{competitive relationship} and \emph{competitive intensity} prediction, and the combination of the coarse-grained and fine-grained competitive features (CF+FF) can further improve the performance in competition prediction.

\vspace{0.3cm}
\makeatletter\def\@captype{figure}\makeatother
\begin{minipage}{0.45\linewidth}
	\centering
	\includegraphics[height=40mm]{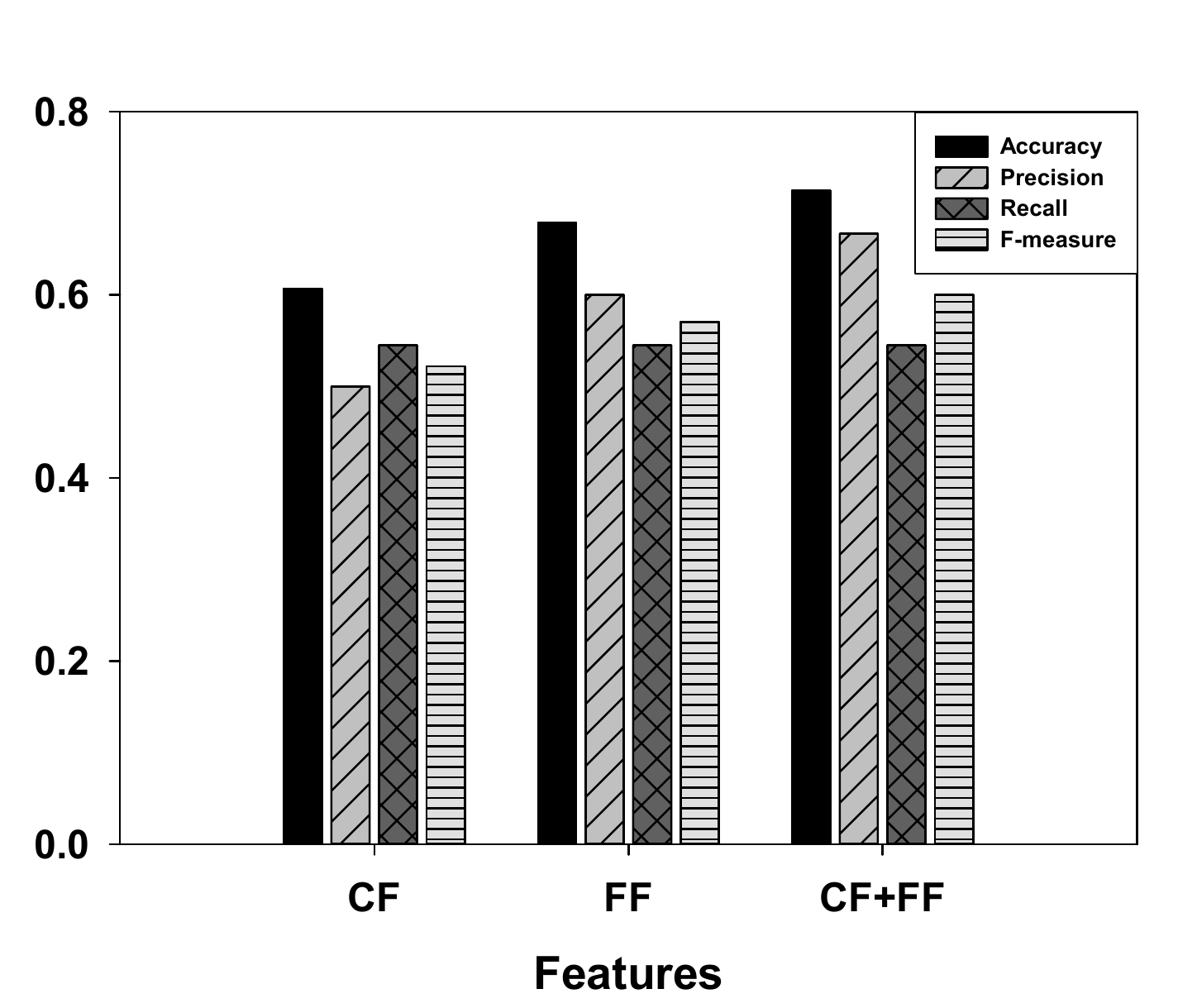}
	\caption{Comparison of Features.}
	\label{fig:features}
\end{minipage}
\makeatletter\def\@captype{table}\makeatother
\begin{minipage}{0.55\linewidth}
	\centering
	\caption{RMSE of Features}
	\label{tab:features}
	\vspace{0.3cm}
	\begin{tabular}{m{1.5cm}<{\centering} m{1.5cm}<{\centering} m{1.5cm}<{\centering}}
		\hline
		CF & FF & CF+FF\\
		\hline
		0.2059 & 0.1980 & 0.1886\\
		\hline
	\end{tabular}
\end{minipage}

\subsubsection{Comparison of Different Data Sources.}
We aim to determine whether the combination of app store data and microblogging data can improve the performance of prediction. Therefore, we compare the AF, MF, and AF+MF, respectively.

Figure~\ref{fig:data_sources} shows the Accuracy, Precision, Recall and F-measure of AF, MF and AF+MF. We can observe that AF outperforms MF, with the Accuracy of 64.3\%, while MF is 60.7\%. This is because that AF constitutes reviews and scores which can reflect users' online experience with the app. Users may report their sentiment or requirement through reviews, and their satisfaction degree through rating scores. It will directly affect the popularity of the app, therefore will affect the popularity contest. In contrast, MF reflects the popularity contest indirectly. Furthermore, the combination of features from app store and microblogging (AF+MF) improves the performance in \emph{competitive relationship} prediction, compared with AF and MF alone.

Table~\ref{tab:data_sources} shows the RMSE of AF, MF and AF+MF. We can observe that AF outperforms MF, because AF will directly affect the popularity of the mobile app, while MF reflects the competition indirectly. Furthermore, the combination of features from app store and microblogging (AF+MF) improves the performance in \emph{competitive intensity} prediction, compared with AF and MF alone.

In summary, AF outperforms MF in both \emph{competitive relationship} and \emph{competitive intensity} prediction, and the combination of features from app store and microblogging (AF+MF) further improve the performance in competition prediction.

\vspace{0.3cm}
\makeatletter\def\@captype{figure}\makeatother
\begin{minipage}{0.45\linewidth}
	\centering
	\includegraphics[height=40mm]{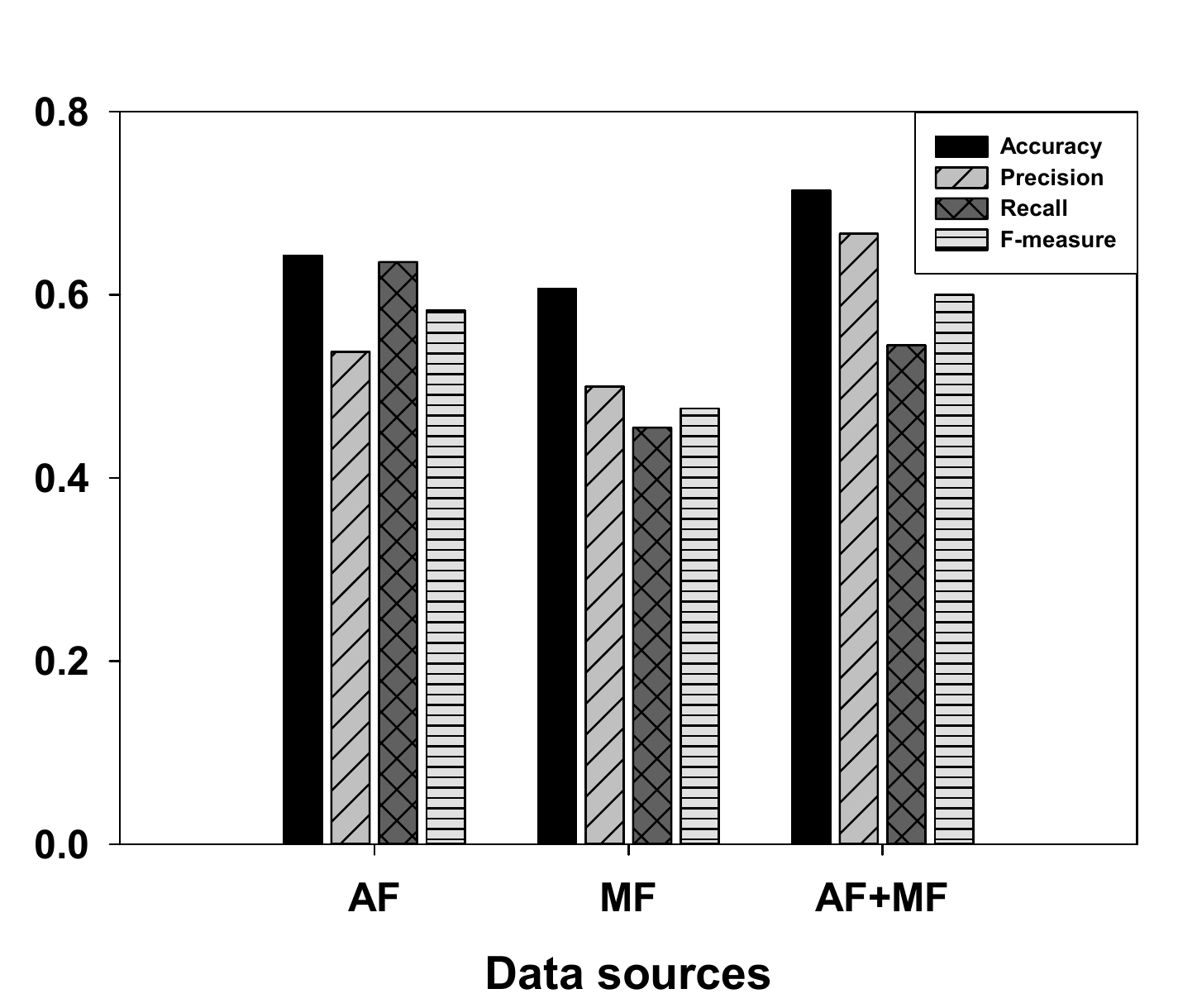}
	\caption{Comparison of Data Sources.}
	\label{fig:data_sources}
\end{minipage}
\makeatletter\def\@captype{table}\makeatother
\begin{minipage}{0.55\linewidth}
	\centering
	\caption{RMSE of Data Sources}
	\label{tab:data_sources}
	\vspace{0.3cm}
	\begin{tabular}{m{1.5cm}<{\centering} m{1.5cm}<{\centering} m{1.5cm}<{\centering}}
		\hline
		AF & MF & AF+MF\\
		\hline
		0.1965 & 0.2062 & 0.1886\\
		\hline
	\end{tabular}
\end{minipage}

\section{Conclusion}
In this paper, we focus on the problem of competitive prediction over Mobike and Ofo. We propose CompetitiveBike to predict the popularity contest between Mobike and Ofo leveraging heterogeneous app store data and microblogging data. Specifically, we first extract features from different perspectives including the inherent descriptive information of apps, users' sentiment, and comparative opinions. With the basic information, we further extract two sets of novel features: coarse-grained and fine-grained competitive features. Finally, we choose the Random Forest algorithm to predict the popularity contest. Moreover, we collect data about two bike-sharing apps from 11 online mobile app stores and Sina Weibo, implement extensive experimental studies, and the results demonstrate the effectiveness of our approach. 

In the future work, we will enrich our problem statement and system framework by learning from the classical economic theories on competitive analysis \cite{bergen2002competitor}, \cite{borodin2005online}. In order to provide competitive analysis for mobile apps, we will view the mobile apps competition as a long-term event, and generate the event storyline \cite{guo2017crowdstory} and present descriptive information regarding popularity contest to enrich the competitive analysis. Besides, we will improve the prediction model by analyzing the couplings \cite{cao2012coupled}, \cite{cao2014behavior} among features and determining their mutual influence. Moever, we will collect more categories of apps to enrich our datasets, and extend the generality of our approach to other apps.

\subsubsection*{Acknowledgments.} This work was partially supported by the National Key R\&D Program of China (No. 2017YFB1001800), and the National Natural Science Foundation of China (No. 61332005, 61772428, 61725205).

\end{document}